\documentclass[aps,prl,preprint,superscriptaddress]{revtex4-2}
\usepackage{graphicx}% Include figure files
\usepackage{dcolumn}% Align table columns on decimal point

\usepackage[draft]{changes}

\usepackage{miller}
\usepackage[version=3]{mhchem}
\usepackage{placeins}
\usepackage{enumitem}
\newcommand{\C}{$^\circ$C}

\preprint{APL}
\begin{document}

\title[TMDJ]{Limits of III-V nanowire growth based on particle dynamics}
	% Force line breaks with \\
\author{Marcus Tornberg}
	\email{marcus.tornberg@ftf.lth.se}
 	\affiliation{Solid State Physics, Lund University,\\Box 118, 22100, Lund, Sweden}
 	\affiliation{NanoLund, Lund University, \\								 Box 124, 22100, Lund, Sweden}
 	
 	\author{Carina B. Maliakkal}%
 	\affiliation{Solid State Physics, Lund University,\\Box 118, 22100, Lund, Sweden}
 	\affiliation{NanoLund, Lund University, \\								 Box 124, 22100, Lund, Sweden}
\author{Daniel Jacobsson}%
	\affiliation{Centre for Analysis and Synthesis/nCHREM, Lund University, \\	 Box 124, 22100, Lund, Sweden}
	\affiliation{NanoLund, Lund University, \\								 Box 124, 22100, Lund, Sweden}
\author{Kimberly A. Dick}%
 	\affiliation{Solid State Physics, Lund University,\\Box 118, 22100, Lund, Sweden}
	\affiliation{Centre for Analysis and Synthesis/nCHREM, Lund University, \\								 Box 124, 22100, Lund, Sweden}
	\affiliation{NanoLund, Lund University, \\								 Box 124, 22100, Lund, Sweden}
\author{Jonas Johansson}
	\affiliation{Solid State Physics, Lund University,\\Box 118, 22100, Lund, Sweden}
	\affiliation{NanoLund, Lund University, \\								 Box 	124, 22100, Lund, Sweden}
             
\begin{abstract}
Crystal growth of semiconductor nanowires from a liquid droplet depends on the stability of this droplet at the liquid-solid interface. By combining in-situ transmission electron microscopy with theoretical analysis of the surface energies involved, we show that truncation of the interface can increase the stability of the droplet, which in turn increases the range of parameters for which successful nanowire growth is possible. In addition to determining the limits of nanowire growth, this approach allows us to experimentally estimate relevant surface energies, such as the GaAs \hkl{11-20} facet.
 
\end{abstract}

%\pacs{ ??? }% PACS, the Physics and Astronomy
                             % Classification System
%\keywords{ ??? }%Use showkeys class option if keyword
\maketitle
 Formation of semiconductor nanowires from a nanoscale liquid droplet by the vapor-liquid-solid process is an important example of complex, multiphase crystal growth. This process allows for the formation of anisotropic crystals with highly precise selectivity for crystal phase\cite{Joyce2010}, composition\cite{Bjork2002,Venkatesan2013}, morphology\cite{Namazi2015,Zou2007} and properties\cite{Im2014}. While the nucleation step and its role in material properties has been investigated in detail\cite{Glas2007,Gamalski2011,Dubrovskii2014}, less attention has been paid to the overall stability of the process and the conditions required for the droplet to remain at the top of the nanowire during growth\cite{Nebolsin2003,Roper2010}. This condition is essential to prevent nanowire kinking and spontaneous change of growth orientation\cite{Schwarz2009,Wang2012}, displacement of the droplet from the nanowire\cite{Tornberg2017, Ghisalberti2019} and failure of the growth process itself\cite{Nebolsin2003}. A fundamental stability criterion for a droplet to remain at the top of a nanowire during its growth has been proposed by Nebol'sin and Shchetinin\cite{Nebolsin2003} based on ex-situ observations and earlier theoretical work \cite{Young1805,Voronkov1974}(see ref \cite{Makkonen2016} for a modern discussion of Young's equation). The wetting properties of the droplet during nanowire growth have subsequently been investigated; theoretically by addressing the droplet stability \cite{Nebolsin2016,Dubrovskii2017, Ghisalberti2019}, and experimentally by focusing on morphology and nucleation\cite{Davidson2007,Wen2011,Jacobsson2016,Kim2018}. Still, the Nebol'sin-Shchetinin stability criterion remains generally accepted, perhaps due to the simplicity of the model.\bigskip

The Nebol'sin-Shchetinin model predicts an upper bound for having a droplet on the top nanowire facet by relating the ratio of the surface energies of the solid and liquid phases in contact with the vapor ($\gamma_{sv}$ and $\gamma_{lv}$) to the wetting angle and tapering of the nanowire\cite{Nebolsin2003}. Although the model is widely accepted, its predictions frequently disagree with experimental observations: for instance growth of self-assisted GaAs\cite{Cirlin2010} and InAs\cite{Dubrovskii2008} nanowires has been extensively reported, although the relevant surface energy ratios in these cases fall outside the predicted stability range ($\gamma_{sv}/\gamma_{lv}$ $\approx$ 2 compared to the maximum ratio of $\sqrt{2}$ for un-tapered nanowires\cite{Nebolsin2003}). Moreover, the model is based on the assumption that the interface between droplet and nanowire is flat, which, according to in-situ experimental results\cite{Oh2010,Gamalski2011,Wen2011,Jacobsson2016}, is not always the case during growth. These in-situ experimental reports instead indicate a dynamic interface that may become truncated during growth, which could be one of the reasons for the mismatch between experimental and theoretical studies.\bigskip

In this letter, we address the stability of the droplet wetting the nanowire top facet during growth by combining in-situ real-time transmission electron microscopy (TEM) observations of GaAs nanowire growth, with a theoretical model that expands on the Nebol'sin-Shchetinin stability criterion to allow the possibility of interface truncation. Given that the truncating facet has been observed to oscillate, in both size and truncation angle\cite{Gamalski2011,Wen2011}, we also consider a lower limit of the surface energy ratio and wetting of a truncated droplet-nanowire interface to evaluate the condition for which a specific truncation would be probable. By measuring the droplet wetting angle in-situ during growth and estimating its surface tension, we demonstrate that the stability range for nanowire growth with a liquid droplet is extended by formation of a truncated interface. \bigskip

\begin{figure}
\includegraphics[scale=1]{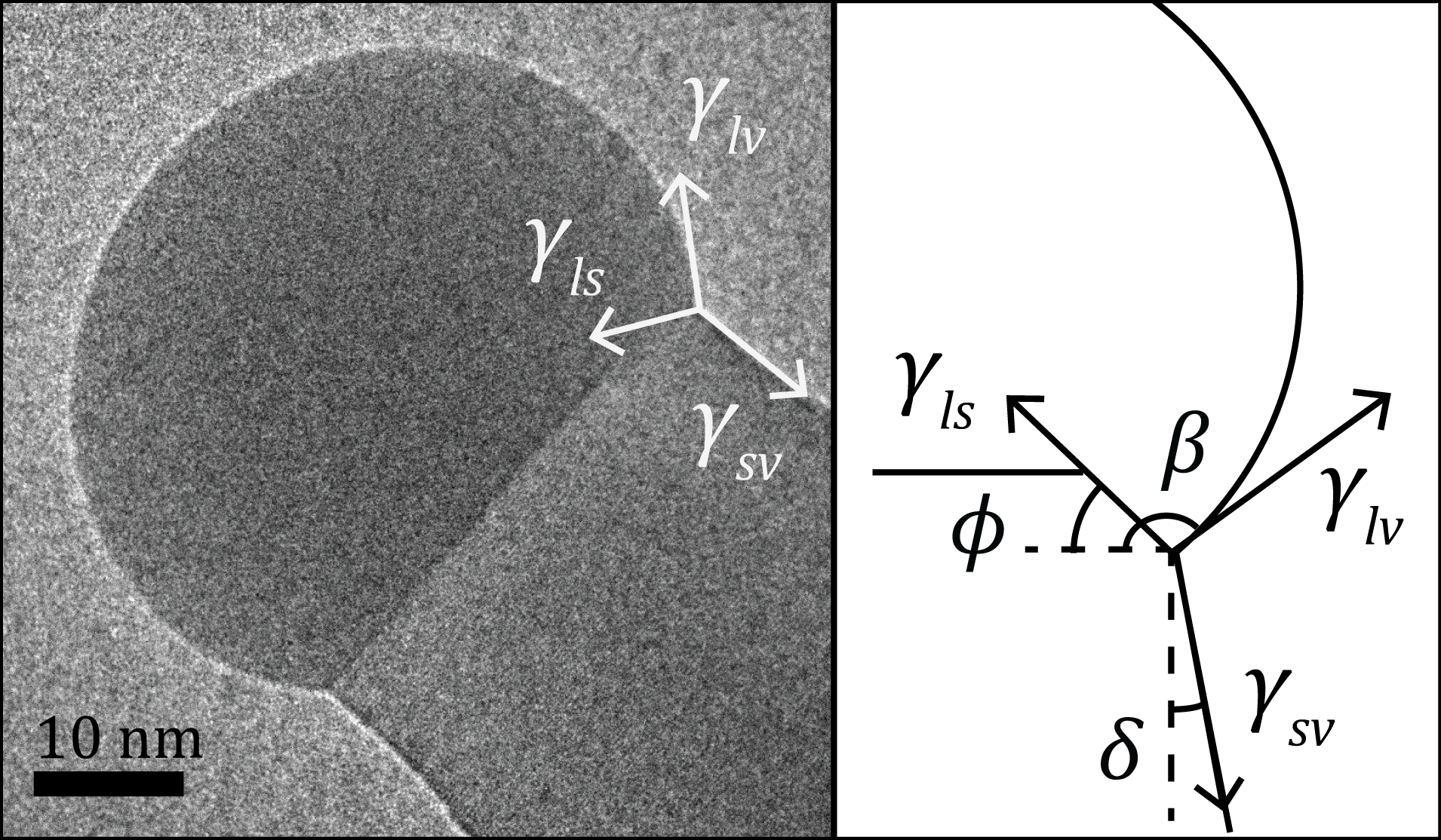}
\caption{The surface forces pulling on a droplet, based on surface energies at the interfaces ($\gamma_{vs}$, $\gamma_{lv}$, $\gamma_{ls}$), are superimposed on a conventional transmission electron micrograph of a Au-droplet on top of a GaAs nanowire. This overview is accompanied by a schematic illustrating the angles used for orienting these forces with respect to each other; taking into account the dependence of truncation ($\phi$), tapering ($\delta$) and wetting angle ($\beta$).}
\label{fig:1overview}
\end{figure}
To introduce the possibility of truncation into the stability condition for having a droplet on the top facet of a nanowire, we introduce further geometrical dependence for the balance of the surface (capillary) forces resulting from the corresponding surface energies, as depicted in figure \ref{fig:1overview}. The figure shows an overview of the droplet-nanowire morphology, depicting tapering, wetting and truncation angles as well as the relevant surface energies.  To balance the surface forces, we consider all solid-vapor, liquid-vapor and liquid-solid interfaces to have the surface energies $\gamma_{sv}$, $\gamma_{lv}$ and $\gamma_{ls}$, respectively. The respective orientation of the interfaces depends on the tapering angle ($\delta$), the wetting angle ($\beta$) and the possible truncation of the top facet ($\phi$). Balancing the horizontal forces laterally at the triple-phase boundary (arrows) in figure \ref{fig:1overview} provides a geometrical relation between the surface energies of the system according to
\begin{equation}
\gamma_{ls}\cos\phi = \gamma_{sv}\sin\delta-\gamma_{lv}\cos\beta
\label{eq:horizonbalance}
\end{equation}
Similarly, the vertical components of the surface forces are evaluated to favor a downward resulting force to study the limits for a droplet to remain stable on the top facet,
\begin{equation}
\gamma_{ls}\sin\phi + \gamma_{lv}\sin\beta < \gamma_{sv}\cos\delta
\label{eq:vertibalance}
\end{equation}
Elimination of $\gamma_{ls}$, using equations \ref{eq:horizonbalance} and \ref{eq:vertibalance}, provides the geometrical condition for the surface energy ratio when the droplet wets part of the nanowire sidewall (i.e. the solid-vapor interface),
 \begin{equation}
\frac{\gamma_{sv}}{\gamma_{lv}} > \frac{\sin\beta \cos\phi-\cos\beta \sin\phi}{\cos\delta \cos\phi -\sin\delta \sin\phi}.
\label{eq:lowerLimit}
\end{equation} 
which reduces to $\gamma_{sv}/\gamma_{lv} > \sin\beta$ for un-tapered nanowires with a flat growth interface ($\delta$=0$^\circ$ and $\phi$=0$^\circ$). Thus equation \ref{eq:lowerLimit} represents a lower limit for the surface energy ratio.\bigskip

\begin{figure}
\includegraphics[scale=1]{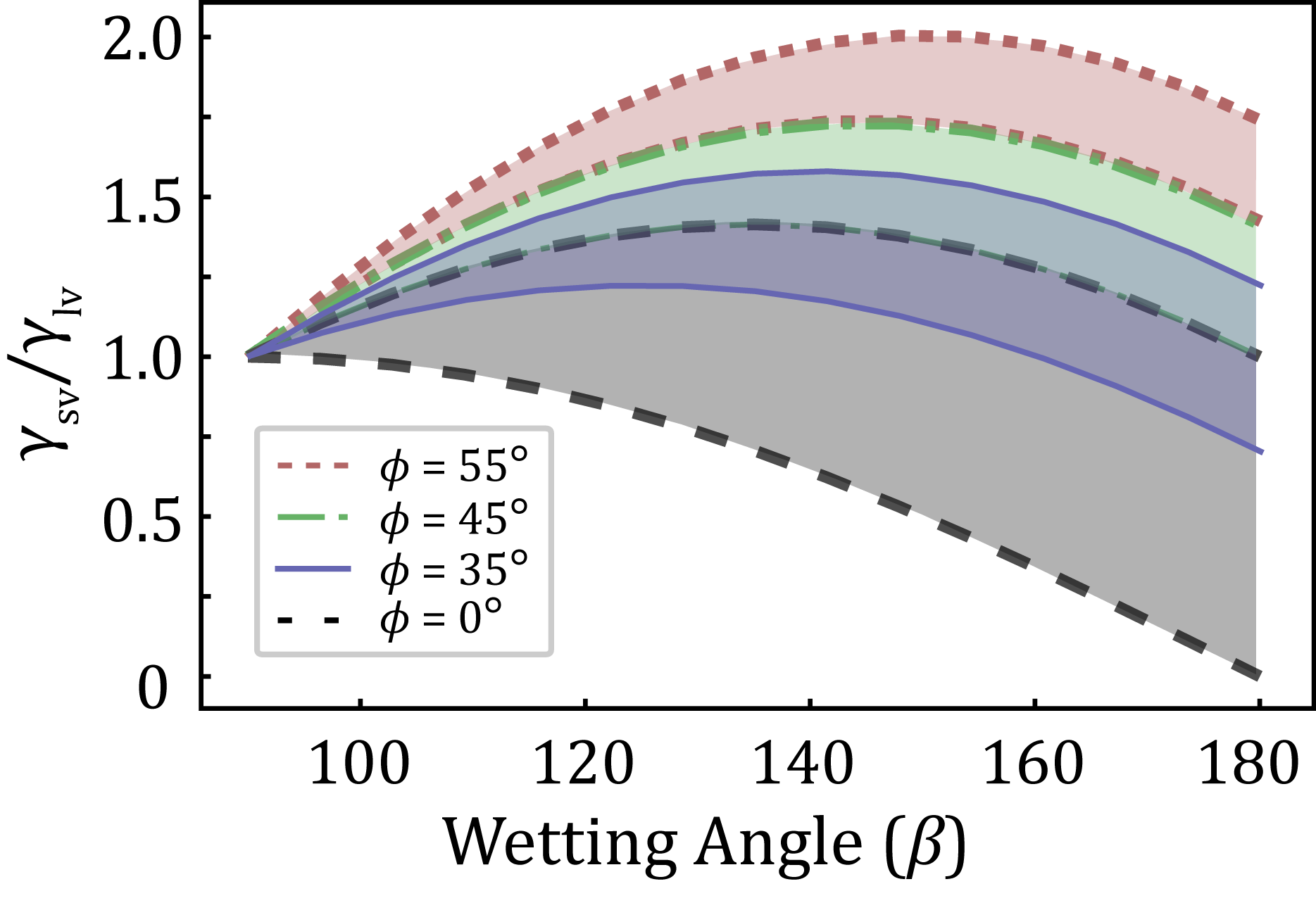}
\caption{A graph over the theoretically predicted an interval for having a droplet on the nanowire top facet ($\delta = 0^\circ $) during growth, bounded by equation \ref{eq:lowerLimit} and \ref{eq:upperLimit} for a truncation angle of 0$^\circ$ (black), 35$^\circ$ (blue), 45$^\circ$ (green) and 55$^\circ$ (red). We observe how the maximum allowed surface energy ratio increases as the truncation angle increases. Note how the lower bound for 45$^\circ$ and 35$^\circ$ overlap with the upper bound for 0$^\circ$ and 45$^\circ$, respectively.}
\label{fig:2cases}
\end{figure}

For the droplet to remain stable on the top facet while having a downward resulting force requires that the resulting force must be directed upwards as soon as the liquid starts to wet the nanowire sidewall. If the resulting force continues to be downward, the droplet would be expected to be displaced from the top facet to the sidewall\cite{Kelrich2016,Tornberg2017}. This results in an upper bound for the surface energy ratio to allow the droplet to remain on the top facet and can be represented by following inequality,

\begin{equation}
\frac{\gamma_{sv}}{\gamma_{lv}} < \frac{\sin\beta \cos\phi - \cos\beta \cos\delta}{\cos\delta \cos\phi - \sin\delta \cos\delta}
\label{eq:upperLimit}
\end{equation}
This ratio reduces to $\sin\beta-\cos\beta$ for un-tapered nanowires with a non-truncated interface between droplet and nanowire, just as in the original model\cite{Nebolsin2003}. The bounds presented, equations \ref{eq:lowerLimit} and \ref{eq:upperLimit}, are drawn in figure \ref{fig:2cases} for un-tapered nanowires with a flat growth interface ($\phi=0^\circ$) and for truncated growth fronts ($\phi>0^\circ$). Here it is evident that the truncation itself extends the stability limit for having a droplet on the top of a nanowire, or a pillar-like structure, allowing higher surface energy ratios than $\sqrt{2}$. \bigskip

\begin{figure}[h]
\includegraphics[scale=1]{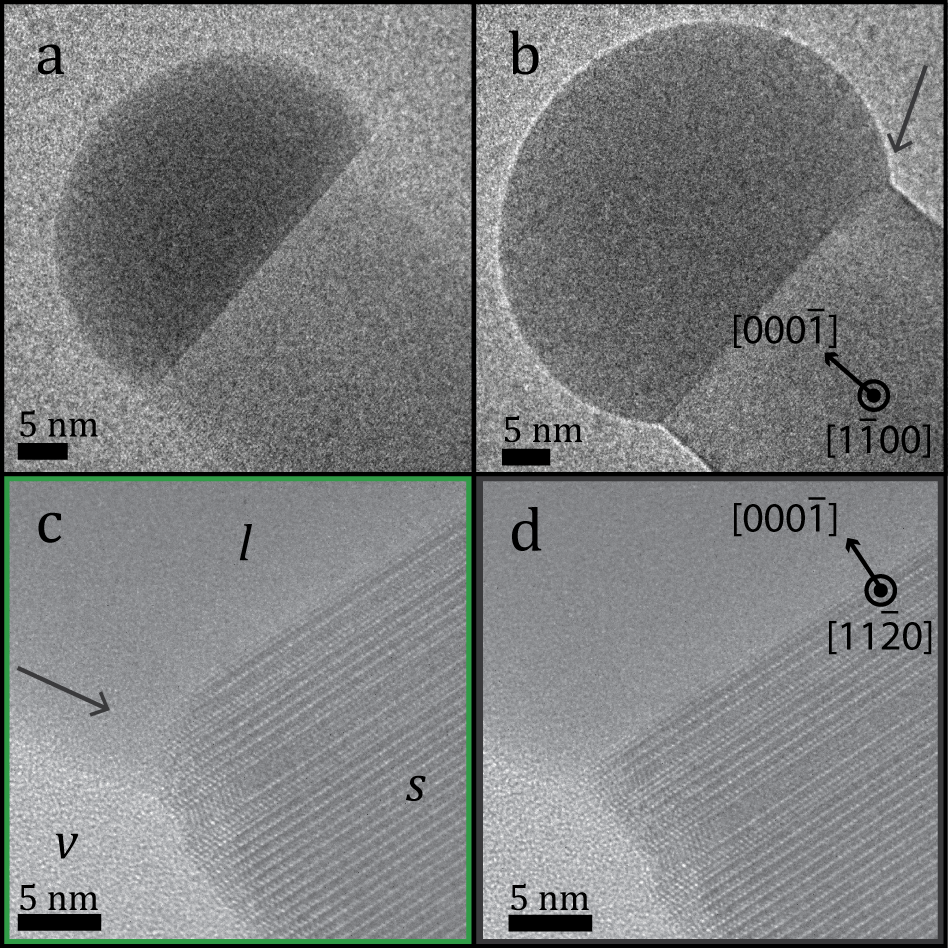}
\caption{As the Ga flow is increased the Au-Ga droplet (darker contrast) is observed to increase its volume (a, b) during the TEM recording of the wurtzite crystal growth. As the droplet was allowed to expand, we observe a truncation of the edge of the interface between the nanowire and the droplet, indicated by arrows (b, c). However, this is not always present but dynamically moves with the droplet and returns to a flat interface from time to time (d).\cite{SupportingInfo}}
\label{fig:3time}
\end{figure}

\begin{figure}[h]
\includegraphics[scale=1]{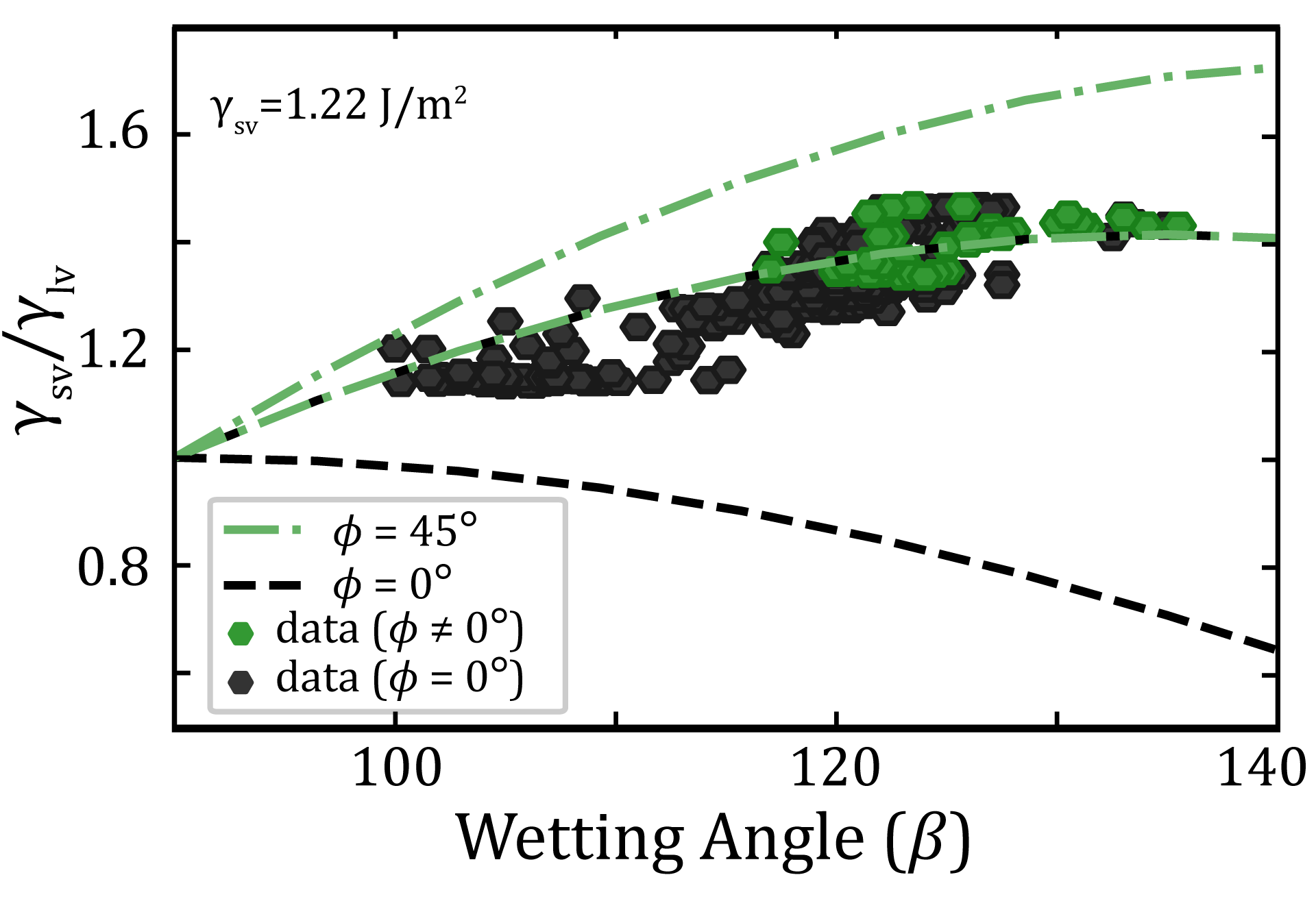}
\caption{The surface energy ratio ($\gamma_{sv}/\gamma_{lv}$) as a function of droplet wetting angle ($\beta$) to the horizontal crystal facet. The dashed lines mark the lower and upper limit for having a droplet wetting the top facet for an un-tapered ($\delta=0^\circ $) nanowire, with (green) and without (black) truncation. The accompanied data (hexagons) are the experimental result from the in-situ microscopy in this letter, measured both when a truncation is present (green) and not (black).}
\label{fig:4data}
\end{figure}

To test the predictions of the model, \hkl(000-1)-oriented Au-assisted wurtzite GaAs nanowires were grown in a transmission electron microscope (TEM) by supplying tri-methyl-gallium (TMGa) and arsine to a \ce{SiN_x} grid, locally heated to 420 \C\cite{SupportingInfo}. When successively increasing the flow of Ga precursor, the size of the Au-Ga droplet was observed to increase as presented in figure \ref{fig:3time}a-b. The volume increase is attributed to Ga accumulation in the droplet, which in turn would lower the droplet surface tension since the Au content remains the same and Ga has a lower surface tension. This allowed us to study several combinations of surface energy ratios and wetting angles, in order to test our model using growth parameters similar to previous reported work on Au-GaAs nanowire growth\cite{Maliakkal2019Feb,Maliakkal2019May}. In order to compare the experimental observations of the droplet to the model, we estimated the volume of the droplet from continuously recorded TEM images with 50~ms exposure time. The estimated change in droplet volume has been shown to provide a good indirect measurement of the change in composition of Au-Ga droplets during nanowire growth\cite{Maliakkal2019Feb}. Based on the volume change, we extract the Ga-concentration in the droplet using a reference quantification done by X-ray energy dispersive spectroscopy\cite{SupportingInfo,CliffLorimer1975} of the same nanowire.
As the size of the droplet increases, we observe truncation of the liquid-solid interface as seen in figure \ref{fig:3time}c. However, this truncation is not always present during the conditions for our growth as shown by the snapshot taken 2 s later, which is presented in figure \ref{fig:3time}d. Based on image recordings, provided as supplementary materials, we observe that the truncation size to change in time, similar to previous reports which have connected it to the droplet supersaturation\cite{Gamalski2011}. In addition, we observe the average truncation angle to vary from 35$^\circ$ to 55$^\circ$ between truncation events. Based on the image recordings during GaAs nanowire growth as the Ga flow into the droplet was successively increased, we measured the wetting ($\beta$) and tapering angles ($\delta$) and estimated the liquid tension based on the droplet volume. These parameters, along with the truncation angle ($\phi$), were used to compare our stability model in figure \ref{fig:2cases} with experimental data. For this comparison, we display the stability regime for a non-truncated interface facet and for the average experimental truncation angle ($45^\circ$) in figure \ref{fig:4data}. \bigskip

The experimental data, included in figure \ref{fig:4data} as data points, have been extracted from measurements of the wetting angle and the estimate of the droplet volume from its two-dimensional projection. Assuming that the recording of the droplet is a two-dimensional projection of a spheroidal cap, and that any added volume to the droplet is pure Ga, allows for an estimation of the droplet composition for each frame of interest. From the composition, the surface tension is estimated by linear extrapolation from the pure species (Au and Ga\cite{Novakovic2006,Hardy1985}), see supporting information for the details on the estimation\cite{SupportingInfo}. Each extracted data point is then related to whether or not a truncation has occurred recently (within 0.5~s), presented as green or black in figure \ref{fig:4data}. From this data we are able to determine an experimental upper limit for the liquid-solid interface ($\gamma_{ls,000\bar{1}}$) as well as an estimate of the solid-vapor surface energy ($\gamma_{vs,10\bar{1}0}$), which will be discussed below. By evaluating equation \ref{eq:horizonbalance} for the wetting angle ($\beta$) and surface tension ($\gamma_{lv}$) for a nanowire without tapering or truncation ($ \phi, \delta = 0^\circ$) we observe that $\gamma_{ls,000\bar{1}}$ does not exceed 0.6~J/m$^2$. Details and limitations are presented as supporting information. In order to estimate the surface energy of the nanowire sidewall ($\gamma_{vs}$) we utilize our observations together with the stability interval presented in this letter (equation \ref{eq:lowerLimit} and\ref{eq:upperLimit}). \bigskip

From the figure, we observe that the experimental observations of non-truncated interfaces (black) correlate with droplets having lower surface energy ratio ($\gamma_{sv}/\gamma_{lv}$) in comparison to most of the cases where truncation is present (green). On the other hand, there is an overlapping region for wetting angles above 120$^\circ$ where both truncated and non-truncated droplet-nanowire interfaces occur for similar surface energy ratios for certain angles. This is reasonable when taking into account that the interface is changing dynamically when forming or removing a truncation, and that the droplet does not change significantly in volume or shape within the 50 ms between each acquired image within the recording. Further, we observed an increased probability of forming a truncation as the particle size increased. The combination of experimental data and the stability model supports the idea that a truncation of the top facet could increase the stability for having a droplet wetting the top of a nanowire. 

In figure \ref{fig:4data}, we have fitted the solid-vapor surface energy of the nanowire side-facet (\hkl{10-10}) to 1.22~J/m$^2$, which is to be compared with existing theoretical calculations using unreconstructed surface (1.3~J/m$^2$\cite{Sibirev2010}) and density functional theory including surface reconstruction and passivation (0.40$<\gamma_{vs, 10\bar{1}0}< $1.06~J/m$^2$\cite{Leitsmann2007,Galicka2008,Pankoke2011}). For this value of the surface energy, we find that most of the data points for the non-truncated interface (black dots) fall below the predicted upper stability limit for this growth with a flat interface, while most of the data for the truncated interface (green dots) fall above this upper limit, and within the stability range for an interface with a truncation of 45$^\circ$. Changes of this fitted surface energy ($\gamma_{sv} $) result in a vertical shift of the experimental data (plotted data), but not the drawn stability limits (lines) as they depend on the geometrical orientation of the capillary forces. Lowering the surface energy by 0.1~J/m$^2$ will shift all data down (0.08~units) and therefore also shift the data related to a truncation into the non-truncated region and vice verse if increased. Fitting optimization and data for a fitted surface energy of 1.22~$\pm$~0.1~J/m$^2$ are provided as supporting material for visual reference\cite{SupportingInfo}. \bigskip

Using the solid-vapor surface energy as a fitting parameter for our data to the model, we are able to compare the experimentally optimum surface energy with the theoretical predictions made with different reconstructions and methods. Our fitted surface energy of the nanowire sidewall of 1.22~J/m$^2$ is close to the theoretical prediction for the energy of the dangling bonds of the unreconstructed \hkl{10-10} GaAs surface (1.3~J/m$^2$)\cite{Sibirev2010}. Under the condition that the liquid-solid interfaces are similar and that the edge energy is neglected, we provide insight to the fundamental crystal surface property by combining theoretical models and experimental data. While this is an experimental estimation, it is important to note that the surface energy is a key factor for the nucleation theory of crystal growth\cite{Wen2011,Glas2007}. To further develop the theory behind crystal growth, it is of importance to narrow the large interval of surface energies predicted by theoretical estimations (ranging from 0.4\cite{Leitsmann2007} to 1.3~J/m$^2$\cite{Sibirev2010}).\bigskip

To conclude, we have theoretically assessed the droplet stability on the top of a nanowire by addressing the possibility of forming a truncation of the droplet-nanowire interface. Experimentally, we have demonstrated a stability increase, allowing for larger ratios between the surface tension and solid surface energy as an effect of forming a truncation. By combining our model with in-situ TEM observations of Au-assisted growth of wurtzite GaAs nanowires, we evaluated the surface energies involved and estimated $\gamma_{sl,000\bar{1}}$ limited by 0.6~J/m$^2$ and $\gamma_{sv, 10\bar{1}0}$ as 1.22~J/m$^2$. This demonstrates that the combination of in-situ growth observations and theoretical models is a powerful means to assess important material parameters for which there are wide variances in theoretical calculations and limited experimental validation.

\subsection*{Acknowledgement}
The authors acknowledge financial support from the Knut and Alice Wallenberg Foundation (KAW), NanoLund, and the Swedish Research Council (VR).
\bibliography{TruncationGaAs_BibTeX}
\end{document}